\documentclass[12pt,epsf]{article}
\usepackage[pctex32]{graphics}
\makeatletter

\@addtoreset{equation}{section}
\newcommand{\be}{\begin{eqnarray}}
\newcommand{\ee}{\end{eqnarray}}
\newcommand{\ba}{\begin{array}}
\newcommand{\ea}{\end{array}}
\newcommand{\nn}{\nonumber}

\makeatletter \@addtoreset{equation}{section} \makeatother

\begin{document}
\vspace{1cm}
\begin{center}
~\\~\\~\\
{\bf  \Large  M2-KK6 System in ABJM Theory : \\
Fuzzy $S^3$ and Wrapped KK6 }
\vspace{1cm}

                      Wung-Hong Huang\\
                       Department of Physics\\
                       National Cheng Kung University\\
                       Tainan, Taiwan\\

\end{center}
\vspace{1cm}
\begin{center}{\bf  \Large ABSTRACT } \end{center}
We study the properties of M2-KK6 (2D membranes - 6D Kaluza-Klein monopole) solution in ABJM membrane theory.   First, we find a new kind of BPS solution which has six coordinates, contrasts to our previous solutions which have four coordinates.  Next,  we argue that, after wrapping 2 sphere the new solution may correspond to the previous solution of four coordinates.  We analyze the properties therein and conclude that M2-branes described in ABJM theory could expand into fuzzy three sphere plus a wrapped 2 sphere near the KK6 core.  Especially, we show in detail how the fuzzy 3-sphere could arise in these solutions and discuss the property of wrapped KK6 and its relation to M5-brane.  We also analyze the fluctuation of the M2-KK6 solution and see that it is U(1) field theory.

\vspace{2cm}
\begin{flushleft}
*E-mail:  whhwung@mail.ncku.edu.tw\\
\end{flushleft}
\newpage
\section{Introduction}
Bagger, Lambert and Gustavsson  (BLG model) [1,2] had proposed a three-dimensional N = 8 superconformal Chern-Simons model  to describe the low energy effective theory of two coincident M2-branes in eleven dimensions [3]. 
Use the Nambu bracket algebras  [4], which are a infinite-dimensional case of Lie n-algebras, the model could describe  infinite  M2 branes [5,6]. 

On the other hand, ABJM [7-10] had proposed an N = 6 Chern-Simons-matter theory with $U(N) \times U(N)$ gauge group, $SO(6)$ R-symmetry and equal but opposite Chern-Simons (CS) levels (k,-k), to capture the dynamics of the low-energy limit of multiple M2-branes on  M-theory orbifold, $C^4/Z_k$.  An important test for the ABJM multiple membranes theory is that it should reproduce the physics of M2-M5 intersections.   The authors in [11,12] had found a kind of  solution and argued that it confirms the property.  However, a detailed investigation in [13] had found that the solution in [11,12] does not described the M5    but merely a single D4-brane interpretation.  Thus the solution in [11,12] is a novel realization of the fuzzy 2-sphere instead of fuzzy 3-sphere.  

As is well-known that, in M-theory there is Kaluza-Klein monopole (KK6) object which is a six dimensional object with coordinate ($t, x_1, x_2, z_1, z_2,y_1, y_2$).   It was argued in [14-15] that  M2-brane, with coordinate ($t,x,z$), can intersect with  KK6 over a 0-brane such that one of the M-2-brane coincides with the isometry direction $(z)$ of the Taub-NUT space [14]
\be
 \left(0|M2,KK6\right)=\Big\{\ba{lrcrrrrrrrr}
t    & x & z&-&-&-&-&-&-&-&-  \\
t  & - & (z)&x_1&x_2&z_1&z_2&y_1&y_2&- &-
\ea
\ee
\\
Thus,  in considering $(0|M2,KK6)$ we need extra 5 dimensional internal spaces to have 6 space dimensions in the world volume of KK6.  This seems to conflict to the conjecture in [11], which said  that the funnel of M2-M5 are the only BPS solution in ABJM theory.   In this paper we will find a new kind of BPS solution which has 6 coordinates, contrasts to our previously found solutions  [16] which have 4 coordinates. This will really enable us to interprete the solution as M2-KK6 system. 

For self-consistence  we first in section 2 review the ABJM model, then we  discuss the BPS solutions with 3 coordinates in [11,12] and solutions with 4  coordinates found in our previous paper [16].   In section 3 we present our new solutions with 6 coordinates.  We also calculate the tension of the solution and show how the new solution could describe the  M2-KK6 (2D membranes with 6D Kaluza-Klein monopole) system.  In section 4  We will argue that, after wrapping 2 sphere the new solution may be corresponding to the  previous solution of 4 coordinates [16].  Using these properties we finally conclude that M2-branes described in ABJM theory could expand into fuzzy three sphere plus a wrapped 2 sphere near the KK6 core.  We follow the nice paper of Nastase, Papageorgakis and  Ramgoolam [13] to see how the fuzzy three sphere geometry could arise in our new solution. We also analyze the fluctuation on the M2-KK6 solution and see that it is U(1) field theory.  We summarize our results in the last section.
\section {ABJM Theory and BPS Solutions}
\subsection {ABJM Theory and BPS Equations}
The ABJM theory is an N = 6 superconformal $U(N)\times U(N)$ Chern-Simons theory of gauge fields $A_\mu$ and $\hat{A}_\mu$ with level (k,-k) coupled to four complex scalars $Y^A$ and four Dirac fermions $\psi_A$, where $A =1, 2, 3, 4$ in the bifundamental representation [7-10], 
\be 
S=\int d^3 x \left[ \frac{k}{4 \pi} \varepsilon^{\mu\nu\lambda} \mathrm{Tr} \left(A_{\mu} \partial _{\nu} A_\lambda + \frac{2i}{3} A_{\mu} A_{\nu} A_{\lambda} 
- \hat{A}_{\mu} \partial_{\nu} \hat{A}_{\lambda} 
- \frac{2i}{3} \hat{A}_{\mu} \hat{A}_{\nu} \hat{A}_{\lambda} \right)\right. \nn\\  
\left. - \mathrm{Tr} D_{\mu} Y_A^{\dagger} D^{\mu} Y^A 
- i \mathrm{Tr} \; \psi^{A \dagger} \gamma^{\mu} D_{\mu} \psi_A
 - V_{\mathrm{bos}} - V_{\mathrm{ferm}} \right]
\ee
with the potentials
\be
V_{bos}&=& -\frac{4 \pi^2 }{3 k^2} \mathrm{Tr} \Big( 
Y^A Y_A^\dagger Y^B Y_B^\dagger Y^C Y_C^\dagger
+ Y_A^\dagger Y^A Y_B^\dagger Y^B Y_C^\dagger Y^C  \nn\\
&&+4 Y^A Y_B^\dagger Y^C Y_A^\dagger Y^B Y_C^\dagger
-6 Y^A Y_B^\dagger Y^B Y_A^\dagger Y^C Y_C^\dagger \Big),\\
V_{ferm}&=& -\frac{2 i \pi }{k}
\mathrm{Tr} \Big( Y_A^\dagger Y^A \psi^{B \dagger} \psi_B
-\psi^{B \dagger} Y^A Y_A^\dagger \psi_B 
-2 Y_A^\dagger Y^B \psi^{A \dagger} \psi_B
+2 \psi^{B \dagger} Y^A Y_B^\dagger \psi_A  \nn\\
&& -\epsilon^{ABCD} Y_A^\dagger \psi_B Y_C^\dagger \psi_D
+\epsilon_{ABCD} Y^A \psi^{B \dagger} Y^C \psi^{D \dagger} 
\Big),
\ee
ABJM model actually has $SU(4)\sim SO(6)$ R-symmetry and $\cal N$=6  supersymmetry.

In considering BPS solution, which have the dependence of only one of the spatial worldvolume coordinate, say $``s"$, the BPS equations of ABJM theory could be obtained by combining the kinetic terms and potential terms in the Hamiltonian and rewriting it as a sum of perfect squares plus some topological terms [12]:   If we denote $Y^A = (Z^1,Z^2,W^1,W^2)$ then the formula used in this paper is [12]
\be H &=&\int dx ds~{tr}(|\partial _{s}W^{\dagger A}+\frac{2\pi }{k}%
(W^{\dagger B}W_{B}W^{\dagger A}-W^{\dagger A}W_{B}W^{\dagger
B}-Z^{B}Z_{B}^{\dagger}W^{\dagger A}+W^{\dagger A}Z_{B}^{\dagger
}Z^{B})|^{2}  \nn \\
&&+|\partial _{s}Z^{A}+\frac{2\pi }{k}(Z^{B}Z_{B}^{\dagger
}Z^{A}-Z^{A}Z_{B}^{\dagger}Z^{B}-W^{\dagger
B}W_{B}Z^{A}+Z^{A}W_{B}W^{\dagger B})|^{2}  \nn \\
&&+\frac{16\pi ^{2}}{k^{2}}|\epsilon _{AC}\epsilon ^{BD}W_{B}Z^{C}W_{D}|^{2}+%
\frac{16\pi ^{2}}{k^{2}}|\epsilon ^{AC}\epsilon _{BD}Z^{B}W_{C}Z^{D}|^{2}) 
\nn \\
&&+\frac{\pi }{k}\int dx^{1}~{tr}(W_{A}W^{\dagger A}W_{B}W^{\dagger
B}-W^{\dagger A}W_{A}W^{\dagger B}W_{B}+2W^{\dagger
A}W_{A}Z^{B}Z_{B}^{\dagger}  \nn \\
&&-2W_{A}W^{\dagger A}Z_{B}^{\dagger}Z^{B}+Z_{A}^{\dagger
}Z^{A}Z_{B}^{\dagger}Z^{B}-Z^{A}Z_{A}^{\dagger}Z^{B}Z_{B}^{\dagger}),
 \ee
in which the last term is topological and doesn't affect the dynamics in the bulk.  A set of BPS equations which minimize the energy in a given topological sector is : 
\be
0&=&\partial _{s}W^{\dagger A}+\frac{2\pi }{k}(W^{\dagger B}W_{B}W^{\dagger A}-W^{\dagger A}W_{B}W^{\dagger B}-Z^{B}Z_{B}^{\dagger}W^{\dagger A}+W^{\dagger A}Z_{B}^{\dagger}Z^{B})\\
0&=&\partial _{s}Z^{A}+\frac{2\pi }{k}(Z^{B}Z_{B}^{\dagger
}Z^{A}-Z^{A}Z_{B}^{\dagger}Z^{B}-W^{\dagger
B}W_{B}Z^{A}+Z^{A}W_{B}W^{\dagger B})  \\
0&=&\epsilon _{AC}\epsilon ^{BD}W_{B}Z^{C}W_{D}=\epsilon ^{AC}\epsilon _{BD}Z^{B}W_{C}Z^{D}
\ee
When the BPS equations are satisfied the topological term in (2.4) gives the energy of the BPS configuration.
\subsection{BPS Solution of 3 Coordinates}
Assumption that $W^A=0$ the BPS equation could be reduced to a  simple form:
\be 
\partial _{s}Z^{A}+\frac{2\pi }{k}(Z^{B}Z_{B}^{\dagger
}Z^{A}-Z^{A}Z_{B}^{\dagger}Z^{B})=0,~~~A,~B=1,2. \ee
Separate the $s$-dependent and independent part by  
\be
Z^{A}=f(s)G^{A},~~~~~f(s)=\sqrt{\frac{k}{4\pi s}},\ee
$G^{A}$s are $N\times N$ matrices and shall satisfy the relation 
\be
G^{A}=G^{B}G_{B}^{\dagger}G^{A}-G^{A}G_{B}^{\dagger}G^{B}.
\ee
This equation is solved in [11,12] by first diagonalizing $G_{1}^{\dagger}~$ using the $ U(N)\times U(N)$ transformations and find that the another matrix $
G_{2}^{\dagger}~$ must be off-diagonal. The $N$ dimensional irreducible solution is 
\be
(G_{1}^{\dagger})_{m,n} =\sqrt{m-1}~\delta _{m,n},~~~(G_{2}^{\dagger
})_{m,n}=\sqrt{N-m}~\delta _{m+1,n}.\ee
Historically, above solution have two kinds interpretations : 

{\it \bf First interpretation} : Authors in [11,12] regarded  $G_1$, $G_2$  as two complex matrix and thus it has four coordinates which can be used to describe M2-M5 system.  They had also calculated the tension of the configuration from the above BPS funnel solution and see that it is consistent with the well known relation between M5-brane and M2-brane tensions : $2\pi T_5=T_2^2$.   To get feeling about the solution we present following forms:

1. The most simple solution of (2.11) is $2\times 2$ matrix
\be X=diag (0,1),~~~Y=diag(0,0),~~~~Z=\left(\ba  {cc} 0&0\\1&0\ea \right)\ee
It represents funnel solution of 2M2-M5 system [11,12].

2. For the $3\times 3$ matrix the solution of (2.11) is  
\be X=diag (0,1,\sqrt2),~~~Y=0,~~~~Z=\left(\ba  {ccc} 0&0&0\\ \sqrt2&0&0\\0&1&0\ea \right)
\ee 
It represents funnel solution of 3M2-M5 system [11,12]. 

{\it \bf Second interpretation} :  However, Nastase, Papageorgakis and Ramgoolam [13] had seen that, as  $G_1^{\dagger}=G_1$ the solution in (2.11) will represent the funnel solution with coordinates $G_1$, $G_2$ and $G_2^{\dagger}$, which in fact has only 3 coordinates and the funnel solution is fuzzy $S^2$ instead of $S^3$ regarded in [11,12].  They had also studied the fluctuations on the solution to convince the property. 
\subsection{BPS Solution of 4 Coordinates}
Denote the matrix notation $Y^A = (X,\tilde Z,Y,W) f(s)$ the another BPS solutions with 4 coordinates found in  our previous paper [16] are those in the case of  $W=0$.  For $N\times N$ matrix  with $N=m+n+1$  the solutions have following non-zero matrix elements:
\be 
X_{i,i}=(0, 0  , ..................... ,  0 , 0  ,  &{\bf 0}&,   1  , \sqrt 2 , ..................,   \sqrt {n-1},  \sqrt n)\\
Y_{i,i}=(\sqrt m , \sqrt {m-1}, ... , \sqrt 2 , 1  ,&{\bf 0}&, 0 , 0 , ...,  0, 0)\\
\tilde Z_{i,i-1}=( 1 ,\sqrt 2  , ... , \sqrt {m-1} , \sqrt {m} &,& \sqrt {n}, \sqrt {n-1}, ..,  2, 1)
\ee
Notice that above solution is irreducible as $\tilde Z$ matrix does not become a double block forms. To get feeling about the solution we present following forms: 
 
For the $4\times 4$ matrix, with N=1+2+1,  an  irreducible solution is 
\be X=diag (0,{\bf0},1,\sqrt 2),~~~Y=diag(1,{\bf0},0,0),~~~~\tilde Z=\left(\ba  {cccc} 0&1&&\\0&\bf 0&0&0\\&\sqrt 2&0&0\\&0&1&0\ea \right)\ee
in which $\tilde Z_{4\times 4}$ looks, more or less,  like as the overlap of $\tilde Z_{2\times 2}$ matrix in (2.12) with $\tilde Z_{3\times 3}$ matrix in (2.13), with overlapped element ${\bf 0}$. {The property of overlapping therein strongly suggests that this is a irreducible solution.}

In [16] we have calculated the tension therein and find an explanation that  the solution (2.14)-(2.15) describes  $N\times N$ M2 membranes expanding  into fuzzy three sphere plus a wrapped 2 sphere near the KK6 core.  This is along the first interpretation in section 2.2, in which we regard each of $X$, $Y$ and $\tilde Z$ as independent complex matrix.  However, if we adopt the second interpretation in section 2.2, then we have only four coordinates  $X$, $Y$ $\tilde Z$ and $\tilde Z^\dag$.
  
  It is the work of this paper to find BPS solution which does has 6 coordinates which really enable us to describe M2-KK6 system.  Especially, we show in detail how the fuzzy 3-sphere could arise in these solutions and discuss the property of wrapped KK6 and its relation to M5-brane.
 
\section{New BPS Solutions}
\subsection{BPS Solution of 6 Coordinates : Pseudo-Direct-Product Solution}
Denote the notation $Y^A = (X,Z,Y,W) f(s)$ for $N\times N$ matrix  with $N=m+n+1$,  the new BPS solutions with 6 coordinates have following non-zero matrix elements: 
\be
X_{i,i}&=&(0~~~,~~~0~~~~~~,...,~~0~,~0~,~~0,~{\bf 0}~,1,\sqrt 2,\sqrt3,...,\sqrt {n-1},\sqrt n)\\
Y_{i,i}&=&i (\sqrt m,\sqrt {m-1},...,\sqrt3,\sqrt2,~1,~{\bf 0}~,0,~~0,~~0,...,~~~~~0~~~,~~~0)\\
Z_{i,i-1}&=&(0~~~,~~~0~~~,0~~...0~~,~~0~,~~~{0}~~~,~\sqrt n,\sqrt {n-1},...,\sqrt3,\sqrt 2,1)\\
W_{i,i+1}&=&i(1,\sqrt 2,\sqrt3,...,\sqrt {m-1},~~\sqrt m~~,~~{0}~~~, 0~~ , 0 ~~~ 0~~ ,... 0~~ ,, ~~ 0)\ee
Notice that above solution is irreducible as Z and W matrix do not become a double block forms. To get feeling about the solution we present following forms:

For the $4\times 4$ matrix, with 4=1+2+1,  an  irreducible solution is 
\be X=diag (0,{\bf0},1,\sqrt 2),~~~~Z=\left(\ba  {cccc} 0&0&&\\0&\bf 0&0&0\\&\sqrt 2&0&0\\&0&1&0\ea \right)\ee
\be Y=i ~diag (1,{\bf0},0,0),~~~~W=i \left(\ba  {cccc} 0&1&&\\0&\bf 0&0&0\\&0&0&0\\&0&0&0\ea \right)\ee 
Let us make important comments on the above solution.
\\

1. Matrix  $Z_{4\times 4}$ looks, like as the overlap of $Z_{3\times 3}$ matrix in (2.13) with $0_{2\times 2}$, with overlapped element ${\bf 0}$.  Also matrix  $W_{4\times 4}$ looks like as the overlap of $Z_{2\times 2}$ matrix in (2.12) with $0_{3\times 3}$, with overlapped element ${\bf 0}$.

2. It is easily to see that the matrix with above $X$ and $Z$ while taking $Y=W=0$ is  an BPS solution.  As it is corresponding  to  3-coordinates BPS solution in section 2.2.  Similarly, matrix with above $Y$ and $W$ while taking $X=Z=0$ is also an BPS solution. 

3.  If we have a solution with $(X,Z,0,0)_{m\times m}$ and another solution with $(0,0,Y,W)_{n\times n}$ then we can combine them to become a direct product  matrix $(X,Z,Y,W)_{(m+n)\times (m+n)}$.  This new matrix is surely  a solution.   But it is reducible from a direct-product matrix solution. 

4.  Although our new solution looks very similar to a direct product solution it is  irreducible however.  The crucial point is that there is an {\bf overlapped element  ``0"} as can be explicitly seen in (3.5) and (3.6). We call this solution as {\bf pseudo-direct-product solution}. Above solution is a pseudo-direct-product solution of two matrix $M_{2\times 2}$ and  $N_{3\times 3}$.
\\

For more clear we present following  irreducible $6\times 6$ matrix solution, with N=2+3+1.  
\be X=diag (0,0,{\bf0},1,\sqrt 2,\sqrt 3),~~~~Z=\left(\ba  {cccccc} 0&0&0&&\\0&0&0&&&\\0&0&\bf 0&0&0&\\&&\sqrt 3&0&0&0\\&&0&\sqrt 2&0&0\\&&0&0&1&0\ea \right)\ee
\be Y=i ~diag (\sqrt 2,1,{\bf0},0,0,0),~~~~W=i \left(\ba  {cccccc} 0&1&0&&&\\0&0&\sqrt 2&&&\\0&0&\bf 0&0&0&0\\&&0&0&0&0\\&&0&0&0&0\\&&0&0&0&0\ea \right)\ee 
This a {\bf pseudo-direct-product solution} of two matrix $M_{3\times 3}$ and  $N_{3\times 4}$.
\subsection{Proof}
  We now prove that the matrix $X$, $Z$, $Y$ and $Z$ in (3.1)-(3.4) is the  solution of BPS equations (2.5)-(2.7). 
\\
$\bullet$ Step 1: 
\\ Use (3.1)-(3.2) we see that $XY=YX=0$ . This implies
\be ~~~~&\Rightarrow&~~ \epsilon ^{BD}W_{B}X W_{D}=YXW-WXY=0\\
&\Rightarrow&~~ \epsilon ^{BD}Z_{B}Y Z_{D}=XYZ-ZYX=0\ee
$\bullet$ Step 2 :\\
Use (3.3) and (3.4) we see that $ZW=WZ=0$. This implies
\be~~~~~~~&\Rightarrow&~~ \epsilon ^{BD}W_{B}Z W_{D}=YZW-WZY=0\\
&\Rightarrow&~~ \epsilon ^{BD}W_{B}W W_{D}=XWZ-ZWX=0
\ee
Thus the BPS equations (2.7) are satisfied.
\\
\\
$\bullet$ Step 3:\\
Use (3.1)-(3.4) we see that $W^{\dagger B}W_{B}X = W^{\dagger
B}W_{B} Z = XW_{B}W^{\dagger B}=ZW_{B}W^{\dagger B}= 0$. Thus the  BPS equation (2.6) is reduced to be 
\be 0&=&\partial _{s}Z^{A}+\frac{2\pi }{k}(Z^{B}Z_{B}^{\dagger
}Z^{A}-Z^{A}Z_{B}^{\dagger}Z^{B}-W^{\dagger
B}W_{B}Z^{A}+Z^{A}W_{B}W^{\dagger B})\nn\\
&=&\partial _{s}Z^{A}+\frac{2\pi }{k}(Z^{B}Z_{B}^{\dagger
}Z^{A}-Z^{A}Z_{B}^{\dagger}Z^{B})
\ee
This equation is just BPS equation in (2.8) and solution of $X$ and $Z$ in (3.1) and (3.3) is just the BPS solution of 3 coordinates in (2.11) plus a direct product of zero matrix. 
\\
\\
$\bullet$ Step 4 :\\
Use (3.1)-(3.4) we see that $Z^{B}Z_{B}^{\dagger}Y^{\dagger}=Z^{B}Z_{B}^{\dagger}W^{\dagger}=Y^{\dagger}Z_{B}^{\dagger}Z^{B}=W^{\dagger}Z_{B}^{\dagger}Z^{B}=0$. Thus the BPS equation (2.5) is reduced to be 
\be
0&=&\partial _{s}W^{\dagger A}+\frac{2\pi }{k}(W^{\dagger B}W_{B}W^{\dagger A}-W^{\dagger A}W_{B}W^{\dagger B}-Z^{B}Z_{B}^{\dagger}W^{\dagger A}+W^{\dagger A}Z_{B}^{\dagger}Z^{B})\nn\\
&=&\partial _{s}W^{\dagger A}+\frac{2\pi }{k}(W^{\dagger B}W_{B}W^{\dagger A}-W^{\dagger A}W_{B}W^{\dagger B})
\ee
This equation is just BPS equation in (2.8) and solution of $Y$ and $W$ in (3.2) and (3.4) is just the BPS solution of 3 coordinates in (2.11) plus a direct product of zero matrix.  Q.E.D.
\subsection{Energy of BPS Solutions and KK6 Tension in ABJM}
The energy of the solution can be evaluated from  (2.4). 
\be E &=&\frac{\pi }{k}\int dx~\mbox{tr}(W_{A}W^{\dagger A}W_{B}W^{\dagger
B}-W^{\dagger A}W_{A}W^{\dagger B}W_{B}+2W^{\dagger
A}W_{A}Z^{B}Z_{B}^{\dagger}  \nn \\
&&-2W_{A}W^{\dagger A}Z_{B}^{\dagger}Z^{B}+Z_{A}^{\dagger
}Z^{A}Z_{B}^{\dagger}Z^{B}-Z^{A}Z_{A}^{\dagger}Z^{B}Z_{B}^{\dagger}) \nn\\
&=&2\int dsdx\mbox{tr}(\partial _{s}Z_{A}^{\dagger}\partial _{s}Z^{A}+\partial _{s}W_{A}^{\dagger}\partial _{s}W^{A})={k\over 8\pi} \Big(m(m+1)+n(n+1)\Big)\int dsdx ~s^{-3}\nn\\
&=&{k\over 8\pi}\{\ba {ccc}
{N^2-1\over2}\int dsdx ~s^{-3},~~~~&& odd~N\\
{N^2\over2}\int dsdx ~s^{-3},~~~~&& even~N\ea
\ee
Note that we have used the BPS equations (2.5) and (2.6) to obtain the second line. In the last line we consider only the lowest energy case among these solution for a fixed value of $N =m+n+1$. Now, introducing the dimension parameter of M2 tension $T_{M2}$ [12] we can define the radius $R$ averaged over the M2 by
\be
R^2= {2{tr (Y^AY_A^{\dagger})}\over N}=\Big\{\ba {cc} {N^2-1\over N}{k\over4\pi T_{M2}}{1\over s},~~&odd~N \\
\\{N}{k\over4\pi T_{M2}}{1\over s},~~&even~N  \ea
\ee
In terms of $R$ the energy becomes
\be E=\{\ba {ccccc}
{T_{M2}^2\over \pi}{N^2-1\over N^2} \int dx dR~{R^3\over k}~2\pi^2&=&{T_{M2}^2\over \pi}{N^2-1\over N^2}\int dx~d^4y~{1\over k},~~&&odd~N \\
{T_{M2}^2\over \pi} \int dxdR~{R^3\over k}~2\pi^2 &=& {T_{M2}^2\over \pi}~\int dx~d^4y~{1\over k},~~~&&even~N 
\ea\ee
To proceed we shall notice that in our solution the values of $X,Y,Z,W, Z^\dag, W^\dag$ are non-zero and we have six coordinates.  However, the radius defined in (3.14) will always produce the $dR~R^3$ in the energy.  This will always produce the volume integration $dx~d^4y$ which, at first sight, could not provide a sufficient space dimension, six, to represent the M2-KK6 system.

    To solve the puzzle let us first notice that  M2-brane can intersect with  KK6  over a 0-brane such that one of the worldvolume with coordinate ($t,x,z$) of the M-2-brane coincides with the isometry direction $``z"$ of the Taub-NUT space [14] (see (1.1)).   In this case M2 is a wrapped configuration [15].  Thus the integrations $\int dx$ in (3.17)  shall be taken over a wrapped value of $\int dx= g_s\ell_s$.  (Note that $R_{11}=g_s\ell_s$) and (3.17) becomes
\\
\be
 E&=&{T_{M2}^2\over \pi}{N^2-1\over N^2}g_s\ell_s\int {d^4y\over k}={T_{M2}^2\over \pi}{N^2-1\over N^2}g_s\ell_s\cdot {1\over 4\pi \ell_s^2}\cdot\int {d^6y\over k}\nn\\
&=&{N^2-1\over N^2}T_{KK6}\int {d^6y\over k},~~~~~odd~N\\
\nn\\
E&=&{T_{M2}^2\over \pi}g_s\ell_s~\int {d^4y\over k}={T_{M2}^2\over \pi}g_s\ell_s\cdot {1\over 4\pi \ell_s^2}\cdot\int {d^6y\over k}\nn\\
&=&T_{KK6}\int {d^6y\over k},~~~~~even~N 
\ee
\\
in which we have added two sphere volume $\int dy^2=\int_{\ell_s} d\Omega_2=4\pi \ell_s^2 $.   This result is just that in [16].

    This means that the KK6 in M2-KK6 system  described in ABJM theory shall be wrapped with 2 sphere with radium $\ell_s$.  Thus M2 membranes could expand into fuzzy $S^3$ plus a wrapped 2 sphere (with radium $\ell_2$) near the KK6 core.   In this interpretation we have a KK6 from M2 in ABJM theory.   
We now detail the property in next section.
\section{Fuzzy $S^3$, Wrapped KK6 and 2M5}
We discuss some properties of above new BPS solution.
\subsection{Fuzzy $S^3$ in Six Coordinates Solution}
In this subsection we investigate the geometry of new solution following the paper of Nastase, Papageorgakis and  Ramgoolam [13]. 

First, we can easily calculate the following $Y_AY^\dag_B$  bilinears, in which $Y_A=(X,Z)$:
\be
(XX^\dag)_{i,i}&=&(0~,...,~0,~{\bf 0}~,1, 2,3,..., {n-1}, n)\\
(ZZ^\dag)_{i,i}&=&(0~,...,~0,~{\bf 0}~,m, m-1,m-2,..., 1)\\
(XZ^\dag)_{i,i+1}&=&(0~,...,~0,~ 0~,\sqrt {m-1}, \sqrt 2 \sqrt {m-2}, \sqrt 3 \sqrt {m-3},..., , \sqrt {m-1})\\
(ZX^\dag)_{i,i-1}&=&(0~,...,~0,~ 0~,\sqrt {m-1}, \sqrt 2 \sqrt {m-2}, \sqrt 3 \sqrt {m-3},..., , \sqrt {m-1})
\ee
If we denote $XX^\dag \equiv J_1^1$, $XZ^\dag \equiv J_1^2$, $ZX^\dag \equiv J_2^1$, $ZZ^\dag \equiv J_2^2$ and define $J_i= (\sigma_i)_\alpha^\beta J^\alpha_\beta$  then we find the following algebra 
\be [J_i, J_j] =2i \epsilon_{ijk} J_k
\ee 
which shows a $SU(2)$ symmetry group, as was checked in [13] for 3 coordinate solution.

Next, we can easily calculate the following $Y_AY^\dag_B$  bilinears, in which $Y_A=(Y,W)$: 
\be
(YY^\dag)_{i,i}&=&(n~,n-1,...........,2,~1,~{\bf 0}~,0,0,..., 0,0)\\
(WW^\dag)_{i,i}&=&(1~,2.............,n-1,~n,~{\bf 0}~,0,0,..., 0,0)\\
(YW^\dag)_{i,i-1}&=&(\sqrt {n-1}~,\sqrt {n-2}.....,~1, 0,0,0,..., 0,0)\\
(WY\dag)_{i,i+1}&=&(\sqrt {n-1}~,\sqrt {n-2}.....,~1, 0,0,0,..., 0,0)
\ee
If we denote $YY^\dag \equiv \tilde J_1^1$, $YW^\dag \equiv \tilde J_1^2$, $WY^\dag \equiv \tilde J_2^1$, $WWZ^\dag \equiv \tilde J_2^2$ and define $\tilde J_i= (\sigma_i)_\alpha^\beta J^\alpha_\beta$  then we find the following algebra 
\be [\tilde J_i, \tilde J_j] =2i \epsilon_{ijk} \tilde J_k
\ee
which constitute another $SU'(2)$ symmetry group as that in (4.5).

Finally, use (4.1)-(4.4) and (4.6)-(4.9) we can check the following relation
\be [J_i, \tilde J_j] =0 \ee
Thus we obtain the $SU(2)\times SU'(2)$ form $Y_AY^\dag_B$  bilinears. As fuzzy $S^3$  has symmetry $SO(4)\sim SU(2)\times SU'(2)$  our new solution thus shows a fuzzy $S^3$  structure.  Let us make following comments to conclude this section. \\

1.  Although the appearance of  symmetry group $SU(2)\times SU'(2)$ is trivial in direct product solution,  it is {\bf not so trivial} in the pseudo-direct-product solution.   The commutation relation (4.11) is  crucial  to make sure the property.

2. We can use bilinears $(X^\dag X)_{i,i}$, $(Z^\dag Z)_{i,i}$,$(X^\dag Z)_{i,i-1}$ and $(Z^\dag X )_{i,i+1}$  to form another ${SU}(2)$ symmetry group, in addition to (4.1)-(4.5).  In a similar way, We can use bilinears $(Y^\dag Y)_{i,i}$, $(W^\dag W)_{i,i}$,$(Y^\dag W)_{i,i-1}$ and $(W^\dag W )_{i,i+1}$  to form another ${SU}(2)$ symmetry group, in addition to (4.6)-(4.10). However, these symmetry merely reflects the bifundamental represent in ABJM theory and is irrelevant to the fuzzy geometry in funnel solution, as detail in [13].
\subsection{Wrapped KK6}
The BPS solution of 4 coordinates in (2.14)-(2.16) found by us in a previous paper [16] also implies tension relation (3.15).  How can we explain this ``coincidence"  ?

First, we see that add matrix $Z$ in (3.5) to $iW^\dag$  in (3.6) will becomes matrix $ \tilde  Z$ in (2.17), i.e. 
\be Z[eq.(3.5)] +iW^\dag[eq.(3.6)] = \tilde  Z[eq.(2.17)]\ee
In fact, this is a general property between the 6-coordinates solution (3.1)-(3.4)  and 4-coordinates solution (2.14)-(2.16).  This property is related to the wrapping  procedure in KK6.   To see the physical reason behind it, let us consider a system with the 2 coordinate (x,y).  We can also use another 2D coordinates (``x+y", ``x-y") to describe this 2D coordinate (x,y).   Now, if we wrap a coordinate ``x-y" then it remains only coordinate ``x+y".   Thus, for a 6 coordinates $(x_1,x_2,x_3,x_4,x_5,x_6)$ system the 4 coordinates $(x_1,x_2,``x_3+x_4",``x_5+x_6")$ system represents the original one while with double wrapped coordinate. 

 In this interpretation, we conclude that our previous solution (2.14)-(2.16) [16] is just a double wrapped of new solution (3.1)-(3.4).

We shall remark that although one can drop any coordinate at hand,  it may be nonsense.  In fact, only if the remained coordinates are also the system solution can we drop some coordinates to describe a physical  system. 

\subsection{Fuzzy $S^3$ in four Coordinates Solution}
   Note that the BPS solution of 3 coordinates has only $SU(2)$ symmetry which corresponds to fuzzy $S^2$,  as detailed in [13].  However, to correspond to  Fuzzy $S^3$ geometry, which is $SO(4)$ symmetry, we need merely 4 coordinate [17] which is less then the 6 coordinates in our new solution. So, what is happen here?  

The reason behind this is that we need to wrap 2 coordinate as mentioned in our interpretation in section 3.3.  Thus, after wrap 2 coordinate we remain 4 coordinate which may corresponds to fuzzy $S^3$.   This then lead to two interesting problems.  

1.  Does the symmetry group in 6-coordinates solution is larger then fuzzy $S^3$?  It seems that it is, but we have not yet found the symmetry group.

2.  Does it appears fuzzy $S^3$ in 4-coordinates solution ? The answer is yes.  This is because that use (4.12) we can ``split" the ``one" $\tilde Z$ coordinate into ``two'' coordinates of $Z$ and $W$.  In this way, as that from section 4.1 we can obtain the fuzzy $S^3$ therefore.\\

   Finally, we shall remark that although one can  ``split"  any coordinate into summation of two arbitrary coordinates, it may be nonsense.  In fact, in our ``splitting" the final two coordinates are also the BPS solution of ABJM theory.  Thus, only if the split coordinates are also the system solution can we study the property of original solution from the split system.  In this interpretation, we can find the hidden symmetry in original system (4-coordinates solution) form the  spilt system (6-coordinates solution with fuzzy $S^3$).
\subsection{Fluctuation}
We now consider the fluctuations around the new solution following the paper of Nastase, Papageorgakis and  Ramgoolam [13]. The investigations of fluctuation on fuzzy $S^3$  are more involved.   The key point is that we first take a small fluctuation around the BPS solution of coordinate $Y_A$. For example, the fluctuation of matrix $Y(s)_\alpha $ around the solution $Y^0_\alpha$   may be expressed as a general form
\be
Y(s)_\alpha = f(s)Y^0_\alpha + r_\alpha
\ee
in which $r_\alpha$ is the fluctuation field and contains U(1) field  [13]. We then expand  the ABJM Lagrangian and identify 
\be [J_i, ]\rightarrow \epsilon_{ijk}x_j\partial_k\ee
in which $J_i$  are defined in section 4.1. Now, after the calculations it is found that the fluctuation on fuzzy $S^2\sim SU(2) $ will show U(1) field [13].  As the case in [13] has only 3 coordinates while our solutions have 6 coordinates it is straightforward to see that the fluctuation on fuzzy $S^3\sim SU(2)\times SU'(2)$ will also show U(1) field while on the KK6 worldvolume.  Thus the field theory KK6 worldvolume is U(1) theory. This is consistent with the analysis in [15].  
\subsection{Wrapped KK6 and 2M5}
Use the  relation $2T_5={T_2^2\over \pi}$ the (3.17) tells us that the 4 coordinates solution can be interpretation as M2-2M5 solution.

  If this is right, then the worldvolume theory of KK6, which  is U(1) field theory,  shall be the  worldvolume theory of 2 M5, which is expected to be non-Abelian self-dual 2 form field theory [18].  Then, they  contradict to each other. 

To explain the contradiction let us see the configuration in (4.15) and remind of a fact that,  when we wrap coordinates $y_1$ and $y_1$ in KK6  the result configuration does not correspond to M5.  
\be
\ba{lrcrrrrrrrrrr}
M2 &:&t    & x & z&-&-&-&-&-&-&-&-  \\
KK6 &:&t  & - & (z)&x_1&x_2&z_1&z_2&y_1&y_2&- &-\\
M5 &: &t  & - & z&x_1&x_2&z_1&z_2&-&-&- &-
\ea
\ee
It is known that, after wrap coordinates $z$ the non-Abelian self-dual 2 form field theory will be reduced to gauge field theory.  Thus, up to the quadratic fluctuation it just give U(1) type Lagrangian, that is described in section 4.4

Note that the correspondence between 4-coordinates solution of M2-2M5 and  6-coordinates solution of M2-KK6 seems to enable us to find  non-Abelian self-dual 2 form field theory by studying fluctuation on M2-KK6 system.  In fact,  the isometry direction in KK6 lead us to wrap a direction on M5, and this will render self-dual 2 form field theory to be one-form gauge field theory. 

In conclusion, we have following summarizations:\\
\\
1.  3-coordinates solution found in [11,12] shows  fuzzy $S^2$ as studied in [13].
\\
2.  Wrapped isometry in 4-coordinates solution (M2-2M5)  = Wrapped 2 sphere in 6-coordinates solution (M2-KK6).  It shows fuzzy $S^3$.
\\
3. In order to obtain (M2-2M5) system from  (M2-KK6) system we shall first wrap 2 sphere in (M2-KK6) system and then ``open" the isometry coordinate ``z".  It shall be emphasized that,  without ``open" the isometry coordinate we could not obtain  (M2-2M5) system.\\

   Finally, the reader may wonder how can ``two" M5 to form ``one'' KK6. Let us explain the ``combination" as following.  
\\
\\
\\
\scalebox{1}{\hspace{3cm}\includegraphics{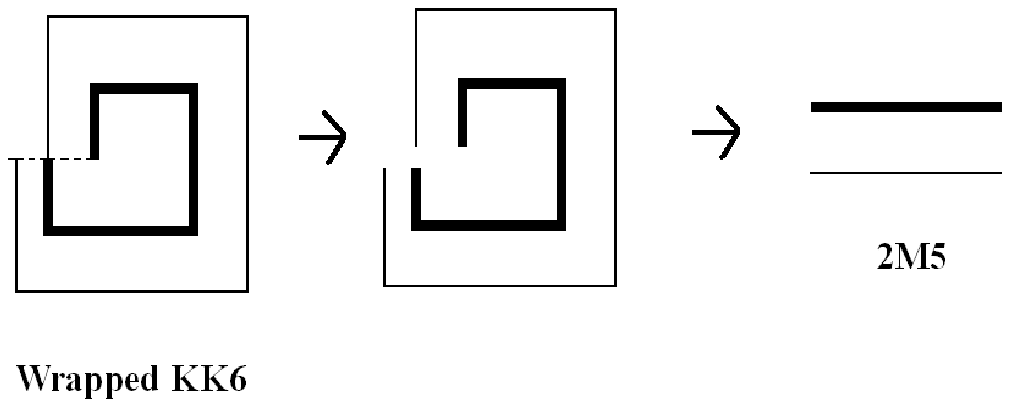}}
\\

{\hspace{2cm} {\it Figure 1. ``One" wrapped KK6 produces ``two" M5.}
\\

  Let us see the figure 1.  We first consider ``one'' KK6 wrapped two times around a  circle, $0<\theta\le 2\pi$ (left diagram).  Next, we ``open" the isometry coordinate and cut it at  $\theta= 0$ (middle diagram).  Finally, we extend the two objects and then obtain ``two" M5  (right diagram).

According this explanation 2M5 are just double free M5-branes while without interaction between them.  Thus, the corresponding worldvolume field theory on KK6 is free U(1) theory.

\section {Discussion}
In this paper we  find a new kind of BPS solution which has six coordinates, contrasts to our previous solutions which have four coordinates [16]. We calculate the tension of the solution and show a possible way to explain our solution as  describing the  M2-KK6 (2D membranes with 6D Kaluza-Klein monopole) systems.  We argue that, after wrapping 2 sphere the new solution may correspond to the previous solution of four coordinates.  Using these properties we conclude that M2-branes described in ABJM theory could expand into fuzzy three sphere plus a wrapped 2 sphere near the KK6 core.  We following the  Nastase, Papageorgakis and  Ramgoolam [13] to see how the fuzzy three sphere geometry could arise in our new solutions.  Especially, we  discuss the property of wrapped KK6 and its relation to M5-brane.  We also analyze the fluctuation on the M2-KK6 solution and see that it is U(1) field theory.
\\
\\
{\bf Acknowledgments} :  This work is supported in part by the Taiwan National Science Council. 
\\
\\
\begin{center} {\bf REFERENCES}\end{center}
\begin{enumerate}
\item J. Bagger and N. Lambert, ``Modeling multiple M2s", Phys. Rev. D 75 (2007) 045020  [hep-th/0611108]; J. Bagger and N. Lambert,``Gauge Symmetry and Supersymmetry of Multiple M2-Branes", Phys. Rev. D 77 (2008) 065008  [arXiv: 0711.0955]; J. Bagger and N. Lambert, ``Comments On Multiple M2-branes", JHEP 0802 (2008) 105  [arXiv: 0712.3738].
\item A. Gustavsson, ``Algebraic structures on parallel M2-branes", Nucl.Phys.B811 (2009) 66  [arXiv:0709.1260]. 
\item M. Van Raamsdonk, ``Comments on the Bagger-Lambert theory and multiple M2-branes", JHEP 0805 (2008) 105 [arXiv:0803.3803]; N. Lambert and D. Tong, ``Membranes on an Orbifold", Phys.Rev.Lett.101 2008 041602 [arXiv:0804.1114];  J. Distler, S. Mukhi, C. Papageorgakis and M. Van Raamsdonk, ``M2-branes on M-folds", JHEP 0805 (2008) 038 [arXiv:0804.1256].
\item Y. Nambu, ``Generalized Hamiltonian dynamics", Phys. Rev. D 7 (1973) 2405; L. Takhtajan, ``On Foundation Of The Generalized Nambu Mechanics, "  Commun. Math. Phys. 160 (1994) 295  [hep-th/9301111]; D. Alekseevsky, P. Guha, ``On Decomposability of Nambu-Poisson Tensor", Acta. Math. Univ. Commenianae 65 (1996) 1 ;  V. T. Filippov, ``n-Lie algebras", Sib. Mat. Zh.,26, No. 6 (1985) 126140; Jose A. de Azcarraga, Jose M. Izquierdo ``n-ary algebras: a review with applications" [arXiv:1005.1028].
\item P.-M. Ho, R.-C. Hou, and Y. Matsuo, ``Lie 3-algebra and multiple M2-branes", JHEP 06 (2008) 020, [arXiv:0804.2110]; P.-M. Ho and Y. Matsuo, ``M5 from M2", JHEP 06 (2008) 105, [arXiv:0804.3629]; P.-M. Ho, Y. Imamura, Y. Matsuo, and S. Shiba, ``M5-brane in three-form flux and multiple M2-branes", JHEP 08 (2008) 014, [arXiv:0805.2898].
\item  Wung-Hong Huang, ``KK6 from M2 in BLG " JHEP 1009 (2010) 109 [arXiv: 1006.4100] 
\item O. Aharony, O. Bergman, D. L. Jafferis and J. Maldacena, ``N=6 superconformal Chern-Simons matter theories, M2-branes and their gravity duals," JHEP 0810 (2008) 091  [arXiv: 0806.1218].
\item O. Aharony, O. Bergman and D. L. Jafferis, ``Fractional M2-branes,"
JHEP 0811(2008)  043  [arXiv: 0807.4924]; [9] K. Hosomichi, K. M. Lee, S. Lee, S. Lee and J. Park, ``N=5,6 Superconformal Chern-Simons Theories and M2-branes on Orbifolds," JHEP 0809 (2008)  002  [arXiv:0806.4977].
\item  M. Benna, I. Klebanov, T. Klose and M. Smedback, ``Superconformal Chern-Simons Theories and AdS(4)/CFT(3) Correspondence,"  JHEP 0809 (2008) 072 (2008) [arXiv:0806.1519].
\item K. Hosomichi, K. M. Lee, S. Lee, S. Lee and J. Park, ``N=5,6 Superconformal Chern-Simons Theories and M2-branes on Orbifolds,"  JHEP0809 (2008) 002 [arXiv:0806.4977]; T. Fujimori, K. Iwasaki, Y. Kobayashi, S. Sasaki, ``Classification of BPS Objects in N=6 Chern-Simons Matter Theory,"  JHEP 1010 (2010) 002  [arXiv:1007.1588]
\item  J. Gomis, D. Rodriguez-Gomez, M. Van Raamsdonk, and H. Verlinde, ``A Massive Study of M2-brane Proposals," JHEP 09 (2008) 113 [arXiv:0807.1074].
\item S. Terashima, ``On M5-branes in N = 6 membrane action", JHEP 0808 (2008) 080 [arXiv:0807.0197]; K. Hanaki and H. Lin H, ``M2-M5 systems in N = 6 Chern Simons theory", JHEP 0909  (2008)  067 [arXiv:0807.2074].
\item H. Nastase, C. Papageorgakis, and S. Ramgoolam, ``The fuzzy S2 structure of M2-M5 systems in ABJM membrane theories," JHEP 05 (2009) 123 [arXiv:0903.3966]
\item  R.D. Sorkin, Phys. Rev. Lett. 51 (1983) 87;\\ D.J. Gross and M. Perry, Nucl. Phys. B226 (1983) 29.
\item E. Bergshoeff, M. de Roo, E. Eyras, B. Janssen and J.P. van der Schaar, ``Intersections involving monopoles and waves in eleven dimensions", Class. Quantum Grav. 14 (1997) 2757, [hep-th/9704120]; E. Bergshoeff,  E. Eyras,  and Y. Lozano, ``The Massive Kaluza-Klein Monopole", Phys.Lett. B430 (1998) 77-86 [hep-th/9802199]; E. Bergshoeff, B. Janssen and T. Ortin, ``Kaluza-Klein Monopoles and Gauged Sigma-Models", Phys. Lett. B410 (1997) 132 [hep-th/9706117];  C.M. Hull, ``Gravitational Duality, Branes and Charges", Nucl. Phys. B509 (1998) 216, [hep-th/9705162];  E. Bergshoeff, J. Gomis, P. K. Townsend, ``M-brane intersections from worldvolume superalgebras," Phys.Lett. B421 (1998) 109 [hep-th/9711043].
\item  Wung-Hong Huang, ``KK6 from M2 in ABJM, " JHEP 1109 (2011) 109 [arXiv: 1006.4100] 
\item  Z. Guralnik, S. Ramgoolam, ``On the Polarization of Unstable D0-Branes into Non-Commutative Odd Spheres, " JHEP 0102 (2001) 032 [hep-th/0101001]; S. Ramgoolam, `` On spherical harmonics for fuzzy spheres in diverse dimensions, " Nucl.Phys. B610 (2001) 461 [hep-th/0105006 ];  S. Ramgoolam, ``Higher dimensional geometries related to fuzzy odd-dimensional spheres, " JHEP 0210 (2002) 064 [hep-th/0207111 ].
\item A. Strominger, ``Open p-branes," Phys. Lett. B 383 (1996) 44 [hep-th/9512059]; E.Witten, ``Five-brane effective action in M theory," J. Geom. Phys. 22, 103 (1997) [hep-th/9610234]; C. Hofman, ``Nonabelian 2-forms," [hep-th/0207017]; C. Papageorgakis and C. Samann, ``The 3-Lie Algebra (2,0) Tensor Multiplet and Equations of Motion on Loop Space," JHEP 1105 (2011) 099 [arXiv:1103.6192]; Pei-Ming Ho, Kuo-Wei Huang and Y. Matsuo, ``A Non-Abelian Self-Dual Gauge Theory in 5+1 Dimensions,"  JHEP  (2011) [arXiv:1104.4040]  
\end{enumerate}
\end{document}